# A novel superhard tungsten nitride predicted by machine-learning accelerated crystal structure searching

Kang Xia[1], Hao Gao[1], Cong Liu[1], Jian Sun[1, *], Hui-Tian Wang[1, †] & Dingyu Xing[1]

[1]National Laboratory of Solid State Microstructures, School of Physics and Collaborative Innovation Center of Advanced Microstructures, Nanjing University, Nanjing, 210093, P. R. China.

**Abstract**

Transition metal nitrides have been suggested to have both high hardness and good thermal stability with large potential application value, but so far stable superhard transition metal nitrides have not been synthesized. Here, with our newly developed machine-learning accelerated crystal structure searching method, we designed a superhard tungsten nitride, *h*-$WN_6$, which can be synthesized at pressure around 65 GPa and quenchable to ambient pressure. This *h*-$WN_6$ is constructed with single-bonded $N_6$ rings and presents ionic-like features, which can be formulated as $W^{2.4+}N_6^{2.4-}$. It has a band gap of 1.6 eV at 0 GPa and exhibits an abnormal gap broadening behavior under pressure. Excitingly, this *h*-$WN_6$ is found to be the hardest among transition metal nitrides known so far (Vickers hardness around 57 GPa) and also has a very high melting temperature (around 1900 K). These predictions support the designing rules and may stimulate future experiments to synthesize superhard material.

Kang Xia and Hao Gao contributed equally to this work. Correspondence and requests for materials should be addressed to J. S. (email: jiansun@nju.edu.cn) or H. T. W. (email: htwang@nju.edu.cn)

Transition-metal nitrides (TMNs) are promising candidates for new ultra-hard materials[1–5], due to their outstanding properties, such as comparable thermodynamic stability to cubic $\gamma$-$Si_3N_4$, high melting points, good chemical inertness, high incompressibility and hardness, as well as their better performance in cutting ferrous metals than diamond[6]. It has been found that strong covalent bonding between nitrogen atoms in TMN structures plays a key role in increasing their elastic stiffness and hardness[7]. To synthesize this kind of compounds, high pressure and high temperature (HPHT) conditions are necessary to overcome the energy barriers of breaking nitrogen molecules and mixing elements. For instance, 4*d*- and 5*d*-transition-metal dinitrides with single-bonded N-N pairs were reported to be synthesized with HPHT method, including $PtN_2$, $IrN_2$, $OsN_2$, and $PdN_2$[8–12]. Their bulk moduli were measured to be comparable to that of diamond. Recently, the $\delta$-MoN resintered at 5–8 GPa and 1400–1800 °C, was reported to be the hardest TMN with superconducting properties, its hardness reaches about 30 GPa[13]. The covalent Mo-N bonded was suggested to enhance the hardness.

Based on first-principles calculations, several superhard TMNs (Vickers hardness >40 GPa) have been proposed. For example, a high nitrogen content $Imm_2$-$ReN_3$ with high hardness of 44.4 GPa, was predicted to be stable under high pressure above 40 GPa[14]. Polyhedral stacking with strong covalent N-N bonds was suggested to remarkably improve the mechanical performance. Another superhard hcp $CrN_2$ was predicted to be energetically stable at pressures above 7 GPa, possessing a hardness of 46 GPa[15]. The superhardness effect was attributed to the strong electron localization into *p*–*d* orbital hybridization, induced by the N-N interstitial pre-compression. On the other hand, nitrogen-rich compounds have been considered as potential high-energy-density materials[16–21]. Recent theoretical predictions found that nitrogen can form rings at high pressure[19–21].

Among transition-metal nitrides, molybdenum and tungsten nitrides were found to possess the highest hardness and to be comparable to those of *c*-BN[6, 22]. Several tungsten nitrides have been successfully synthesized, for instance, rock-salt WN and hexagonal WN ($\delta$-WN)[23], hexagonal and rhombohedral $W_2N_3$, and cubic $W_3N_4$[22]. On the theoretical side, several other W-N compounds with large elastic properties were predicted[24–28]. For instance, the NbO-WN (space group (S.G.) $Pm\bar{3}m$), NiAs-WN (S.G. $P\bar{6}m2$) and $MoS_2$ type $W_2N$ compounds were predicted to possess relatively large bulk ($B$ >300 GPa) and shear ($G$ >200 GPa) moduli[25, 27]. A hexagonal $P\bar{6}m2$ of $WN_2$ was calculated to possess high hardness of 36.6 GPa[24]. Another $W_5N_6$ compound was proposed to have hardness of 28 GPa[28], comparable to that of $\alpha$-$SiO_2$[29]. Very recently, the extraordinary strain stiffening in a *hP6*-$WN_2$ structure was reported to remarkably enhance the indentation strength, which exceeds the threshold of superhard materials (40 GPa)[30].

We can infer from the aforementioned examples that the covalent N-related bonds, especially the N-N single bond, have an important influence on the stiffness and hardness of TMNs. Meanwhile, the isotropy or homogeneity in the direction of covalent bonds largely affects the shear modulus of materials, and thus also influences their hardness. Nice examples for this can be found in ultra-incompressible transition-metal borides (bulk modulus >300 GPa)[31–34]. On the other hand, the metallicity of materials will largely reduce their hardness[35–37]. From the aforementioned observations, three clues seem to be useful for designing hybrid superhard materials containing transition metal and light elements as follows. 1) The candidate should be thermodynamically stable at high pressure and also dynamically stable at ambient pressure, allow us to be able to synthesize it eventually. 2) The good candidate must have a band gap, that is to say, it must be a non-metal. 3) The ratio of light element relative to transition metal atoms must be large enough so that the light element atoms can form strong covalent bonds or even networks and frameworks. These clues inspire us to explore special N-related networks in TMNs, such as rings and even cages, where the short and strongly directional N-N covalent bonds should lead to the super-hardness.

In this work, to perform extensive structure search efficiently, we develop a machine-learning accelerated crystal structure prediction method by combining *ab initio* calculations and Bayesian optimization. Taking W-N binary system as a test for this brand new method, we investigate its phase diagram over a wide pressure range of 0–100 GPa. New ground-state and high-pressure phases at ratios of 1:1 and 1:6 are predicted. Interestingly, the newly found $WN_6$ is a superhard material containing $N_6$ rings, its Vickers hardness is evaluated to be around 57 GPa. The novel $N_6$ rings in $WN_6$ are found to be essential for the exotic electronic structures and excellent mechanical properties.

## Results

Using our machine-learning accelerated first-principles crystal structure prediction method (Fig. 1), a series of convex hulls are obtained at pressures of 0, 20, 50 and 100 GPa (Supplementary Fig. 1). Two new W-N compounds at ratios of 1:1 and 1:6 (W:N) are found. The pressure-concentration diagram of stable W-N compounds is plotted in Fig. 2a. The newly predicted $WN_6$ ($R\bar{3}m$, we named as *h*-$WN_6$) are labeled by blue and italic font (Supplementary Fig. 1), it emerges at pressure of 100 GPa and is energetically favorable over a pressure range of 65–100 GPa (Fig. 2a). The enthalpy of another new W-N phase (space group: $P\bar{6}m2$) with 1:1 ratio (*h*-WN) is very close to that of the NiAs-WN structure[25] in the pressure range of 20–45 GPa.

The formation enthalpy of N-rich *h*-$WN_6$ becomes negative at high pressures >31 GPa (Fig. 2b), which indicates its possible formation from elemental tungsten and nitrogen phases. Therefore, high-pressure synthesis of *h*-$WN_6$ can be achieved in the reactions of WN + 5N→ $WN_6$ (green solid line) and $WN_2$ + 4N → $WN_6$ (blue solid line), at high pressures >50 and 65 GPa respectively (Fig. 2d). The formation enthalpy of $P\bar{6}m2$ WN is found to be negative and decrease monotonically with increasing pressure (Fig. 2b). At zero temperature, $P\bar{6}m2$ phase WN predicted in this work becomes more favorable than the NbO-WN structure[27] at pressure of 20 GPa (Fig. 2c). With further increasing the pressure up to 45 GPa, WN transforms from the $P\bar{6}m2$ phase to NiAs-WN structure[25]. The enthalpy of $P\bar{6}m2$ WN structure is slightly lower than that of NiAs-WN at pressures of 20–45 GPa (about 4 meV/atom at 30 GPa as shown in the inset of Fig. 2c), which makes them more or less overlap with each other in the convex-hull diagram. Thus the $P\bar{6}m2$ WN structure predicted in this work fills in a gap in the pressure range of 20–45 GPa in the previous reported phase diagram of WN[28].

The crystal structures of $P\bar{6}m2$ WN and $R\bar{3}m$ $WN_6$ are shown in Fig. 3c and Figs. 3a and b, respectively. Their structure details are listed in Supplementary Table 1. The $P\bar{6}m2$ WN can be constructed from the NiAs-WN unit cells[25] with *AB* stacking along the [001] direction. Interestingly, the $R\bar{3}m$ $WN_6$ is composed of tungsten atoms and armchair-like $N_6$ (*ac*-$N_6$) rings (Figs. 3a and b). Viewed from the [100] direction, the $WN_6$ phase also looks like a sandwich structure. The bond distance between tungsten atom and its nearest neighboring nitrogen atoms ($d_{W_N}$) is about 2.27 Å. The distance between the N atoms ($d_{N N}$) in the $N_6$ ring is around 1.43 Å, which is close to the bond length of $N_2$ pair in $PtN_2$[9]. The strong covalent bonding $N_2$ pair in $PtN_2$ has been found to be beneficial to stabilize the crystal structure and enhance the high elastic moduli of $PtN_2$[51]. This inspires us to study the bonding nature of *ac*-$N_6$ in *h*-$WN_6$.

We have calculated the electron localization function (ELF) of *h*-$WN_6$ for bonding analysis. The contour plot of three-dimension ELF is projected in the (001) plane and cutting through the center of a $N_6$ ring (Fig. 3d). Valence electrons of W and N atoms are obviously found to localize along W-N and N-N bonding directions. ELF contour cutting through three neighboring N atoms (yellow marks) in a $N_6$ ring and along the (110) plane are presented in Figs. 3e and 3f, respectively. The high localization for the N-N bonds and the lone pairs on the nitrogen atoms can be seen clearly. To quantify the charge transfer between W and N atoms in the $WN_6$ cells, Bader's theory of atoms-in-molecules is employed[52]. The Bader charge analysis reveals that the total charge from one W atom to one $N_6$ ring is around 2.4*e*. This indicates that the *h*-$WN_6$ phase predicted here possesses some kind of ionic-like feature. This type of charge transfer and ionic-like feature in noble metal nitrides has been suggested to be one of the major sources of their huge bulk moduli[53].

To discuss the dynamical stability of the $WN_6$ phase, phonon dispersions are calculated at 0 GPa (Fig. 4a), where there is no any imaginary frequency found. This suggests that this high-pressure $R\bar{3}m$ phase is metastable and can be recovered to ambient pressure, which may be useful in real applications. Additionally, the energetically stable $P\bar{6}m2$ WN, is also dynamically stable judged from phonon spectra (Supplementary Fig. 2a).

This *h*-$WN_6$ is estimated to be a semiconductor with small indirect band gap ($E_g$) by the electronic structures calculation (the left panel of Fig. 3g). $E_g$ is evaluated to be around 0.9 eV at 0 GPa, which changes very little with spin-orbit coupling (supplementary Fig. 7). The valence band maximum and conduction band minimum are located at *Z* and *L* in the first Brillouin Zone, respectively. Results of partial DOS (the right panel of Fig. 3g) reveal that W-5*d* and N-2*p* orbits make main contributions to the conduction bands in the energy interval of 1.0–4.0 eV and to the valence bands of -3.0–0 eV, respectively. The indirect band gap is closely related to W-5*d* occupied and N-2*p* bonding states. The W-6*s* electrons almost have no contribution to the DOS around the Fermi level, which also indicates that they may transfer into localized states, as we see from the charge analysis.

The energy gap of $WN_6$ under ambient pressure is evaluated to be 1.58 eV from the hybrid HSE06 functional, which is around 0.68 eV larger than that calculated by PBE functional. This gap monotonously increases to 2.40 eV with increasing pressure to 200 GPa (Fig. 3h), which behaves in the opposite way, compared with that of usual semiconductors, where the band gap usually becomes narrow upon compression. To get a clear understanding of this odd band-gap behavior of *h*-$WN_6$, we further study the responses of equilibrium volume (V) and atomic distances to external pressures, as depicted in Figs. 3h and 3i, respectively. Our calculations show that $Eg$ of $WN_6$ has a linearly dependence on 1/V (Fig. 3i). And $d_{W_N}$ decreases more quickly than $d_{N_N}$ with compressed volume, which means that the N-N bonds in the $N_6$ rings are very strong compared to the W-N interactions. With the formation of the $N_6$ ring, there are still three valence electrons left for one nitrogen atom. Therefore, some electrons transfer from tungsten atom to nitrogen atom, forming two lone pairs. There is rather strong repulsion between the lone pairs, and between them and other electronic states, which opens a gap in this compound. With compression, the repulsions get even stronger when the volume gets smaller, making the energy gap even bigger.

We further study the elastic properties of *h*-$WN_6$ and *h*-WN. The calculated elastic constants are listed in Table 1, compared with the experimental results of $\delta$-MoN[13]. The calculated bulk modulus $B$ and shear modulus $G$ of *h*-$WN_6$ are 302.7 GPa and 315.7 GPa respectively. Although its bulk modulus is not the highest in transition metal nitride, the high values of both $G$ and Pugh modulus ratio ($G/B$ >1.0) indicate that the *h*-$WN_6$ may have high Vickers hardness. The hardness value of $WN_6$ is estimated by semi-empirical models and exact strain-stress calculations. Using the methods of Chen *et al.*[54] and Tian *et al.*[36], the Vickers hardness of $WN_6$ is estimated to be about 57.9 GPa and 56.8 GPa, respectively, which is much harder than the previously known hardest experimental synthesized transition metal nitride $\delta$-MoN ($Hv$ = 30 GPa)[13] and also the theoretically predicted *hP6*-$WN_2$ ($Hv$ = 46.7 GPa)[30]. Therefore, *h*-$WN_6$ is a potential superhard material, hitting the highest record in the Vickers hardness of transition metal nitride up to now. To cross check the stress-strain responses to the atomic deformation similar to the process in the experimental nano indentation hardness test, we calculate the Vickers indentation shear strength (the centerline-to-face angle equals 68 degree) of *h*-$WN_6$ with the method from Ref. 55 and 30. As shown in Fig. 4b, its peak stresses under ideal tensile strains along the [001] direction is found to be weaker than that along the [$\bar{1}$10] and [110] directions. Thus we evaluate the stresses in the easy cleavage plane of (001), under the Vickers indentation shear strains along three high-symmetry directions [$\bar{1}$10], [110] and [210]. As shown in Fig. 4c, we obtain a lowest Vickers indentation shear strength of around 46 GPa at a peak-stress strain of 0.10 for the (001) plane shearing in the [210] direction. The pure shear strength[56] shown in the Supplementary Fig. 5 is slightly smaller than the Vickers indentation shear strength, which is similar to cases in other systems[55, 30]. However, the Vickers indentation shear strength usually has a much better agreement with the experimentally measured Vickers hardness than the pure shear strength[55, 30].

TMNs are widely studied as hard materials not only because of their outstanding mechanical property, but also for their thermal stability and high melting points[6]. We estimated the melting point of superhard *h*-$WN_6$ by employing the $Z$ method[57] (Fig. 4e). The melting temperature is evaluated to be ~1900 K, by the coexistence of solid and liquid phase in two $Z$ curves. Thus the melting temperature of *h*-$WN_6$ structure is considerably higher compared to other nitrides[6]. We also cross check its thermal stability by performing *ab initio* molecular dynamics (*AIMD*) simulations (Fig. 4 and Supplementary Fig. 3). During the entire *AIMD* simulations running for more than 36 picoseconds using *NpT* ensemble[58] at temperature of around 2273 K, the *h*-$WN_6$ structure stays intact and the covalent N-N bonds in *ac*-$N_6$ rings are not broken. The statistically averaged closest N-N bond length is around 1.44 Å (Fig. 4d). These simulations suggest that the $N_6$ rings are kinetically quite stable and can be successfully preserved at ambient pressure and high temperatures.

Since hexagonal $\delta$-MoN was proposed to be very hard[2, 13], we replace the W atom of *h*-$WN_6$ with Mo atom, and find out the synthesis of an isomorphic *h*-$MoN_6$ is also possible at around 95 GPa (Supplementary Fig. 2b and Supplementary Fig. 4). The Vickers hardness of $h$-$MoN_6$ is calculated to be 50.6 GPa, as listed in Table I, which indicates that $h$-$MoN_6$ is another superhard material candidate.

In summary, we developed a new machine-learning accelerated methodology for crystal structure searching based on Bayesian Optimization and *ab initio* calculations. Three guiding factors seem to be important for designing hybrid superhard compounds with transition metal and light elements: the structural stability, the non-metallicity, and a large ratio of light elements. As a test case for both our method and these guiding rules, a systematic search for the stable phases in W-N system has been performed over a pressure range of 0–100 GPa. Two new tungsten nitrides ($P\bar{6}m2$ WN and $R\bar{3}m$ $WN_6$) are predicted to be stable under high pressures and metastable at ambient conditions. Interestingly, the $R\bar{3}m$ $WN_6$ contains armchair-like $N_6$ rings with pure N-N single bonds. There is considerable charge transfer between the tungsten atoms and the $N_6$ rings in this compound, making it exhibit some ionic features. Different to the usual semiconductors, the band gap of this compound has an abnormal broadening behavior under pressure, which mainly due to the repulsion between the lone pairs in the $N_6$ rings. Even more excitingly, $h$-$WN_6$ is estimated to possess the Vickers hardness of 57 GPa by microscopic hardness models. This value sets the highest hardness record of transition metal nitrides. Its superhardness is cross checked by Vickers indentation shear stiffening calculations. This superhard $WN_6$ structure also has very good thermal stability with a high melting point of ~1900 K. We believe these predictions will stimulate future experiments to synthesize this superhard material with interesting electronic properties.

## Methods

### Machine-learning accelerated crystal structure searching

Recently, many methods have been developed to search or predict crystal structures at ambient or extreme conditions[38–47]. The common goal of these methods is to find the global and/or local minima of the free energy surface. Many of theoretical predictions have been verified by experiments, which validates these methods. However, the crystal structure searching process based on *ab initio* calculations are expensive, and the most time-consuming part is the total energy calculations for each crystal structure. How to raise the efficiency (to accelerate predictions) is a significant challenge. Here we proposed and implemented a machine-learning accelerated crystal structure prediction method based on Bayesian optimization[48] to improve the search efficiency and diversity. The complete algorithm process is depicted in the left panel of Fig. 1. Mutation and heredity operators are used to generate new structures within an evolutionary fashion. The energies for the structures can be predicted by a Gaussian Process model, which is one of the major branches of the machine-learning algorithm[49, 50]. An acquisition function, $F(x) = \mu(x) + k\sigma(x)$, is used to select the structures for the next generation, where $\mu(x)$ and $\sigma(x)$ represent the mean and standard deviations of the predictive distribution at the $x$ point in the structural space, respectively. The maxima of this function are the points with both higher uncertainties and better prediction values, which is a trade-off between the exploration (wide range exploration on the energy surface) and exploitation (careful search near the local minima of the energy surface), as one can see from the right panel of Fig. 1. More details about the method can be found in the Supplementary Information.

**Computational codes**

We used the VASP code[59] to perform the structure optimizations and enthalpy calculations. The Perdew-Burke-Ernzerh of potentials were applied within the generalized gradient approximation (GGA-PBE)[60]. The projector-augmented wave (PAW) method was adopted[61]. The structures were relaxed at a high level of accuracy, consisting of a kinetic energy cutoff of 1050 eV, using a $k$-mesh of spacing $2\pi \times 0.03$ $Å^{-1}$ in the Brillouin Zone. Electronic localization functions (ELF) calculated by VASP were displayed by the Visualization for Electronic and STructural Analysis (VESTA)[62]. Electronic band structures and partial densities of states were computed by the WIEN2k code[63]. The hybrid Heyd-Scuseria-Ernzerhof functional (HSE06)[64, 65] was also used to achieve accurate electronic band structures. Phonon modes and frequencies of the stable structures were calculated by the PHONOPY code[66], combined with the VASP code. Elastic properties were computed by VASP, and the bulk and shear modulus were calculated based on Voigt averaging[67]. And the Vickers hardness was computed by model of Chen *et al.*[54] and Tian *et al.*[36]. The melting point and thermal stability of $h$-$WN_6$ are studied by *ab initio* molecular dynamics (AIMD) simulations in the *NVE*[57] and *NpT* ensembles[58] respectively. All MD simulations are performed for the $\sqrt{2}\times 2\sqrt{2}\times 3$ supercell of $WN_6$ unit cell, containing 252 atoms. The data points in *Z*-method[57] calculations dotted in two *Z* curves, are obtained by statistics over the last 2 ps in every trajectory. The simulation runs to 36 ps with a time step of 1 fs using the *NpT* ensemble with a Langevin thermostat.

## Acknowledgements

We are grateful for financial support from the MOST of China (Grants No. 2016YFA0300404 and No. 2015CB921202), the NSFC (Grants No. 51372112 and No. 11574133), the NSF of Jiangsu Province (Grant No. BK20150012), the Fundamental Research Funds for the Central Universities, and the Special Program for Applied Research on Super Computation of the NSFC-Guangdong Joint Fund (the second phase). Some of the calculations were performed on the supercomputer in the High Performance Computing Center of Nanjing University and "Tianhe-2" at NSCC-Guangzhou.

## Author contributions

J. S. designed and supervised the project. K. X. designed the structure, performed *ab initio* calculations; H. G. and J. S. developed the new machine-learning accelerated crystal structure searching method; C. L. analyzed the bond lengths and structure in *AIMD* simulations, performed *Z*-method calculations; K. X., J. S., H. T. W. and D. Y. X. wrote the paper. All authors discussed the results and commented on the manuscript.

## Additional information

Supplementary Information accompanies this paper at doi:xxxx.

## Figures and Table

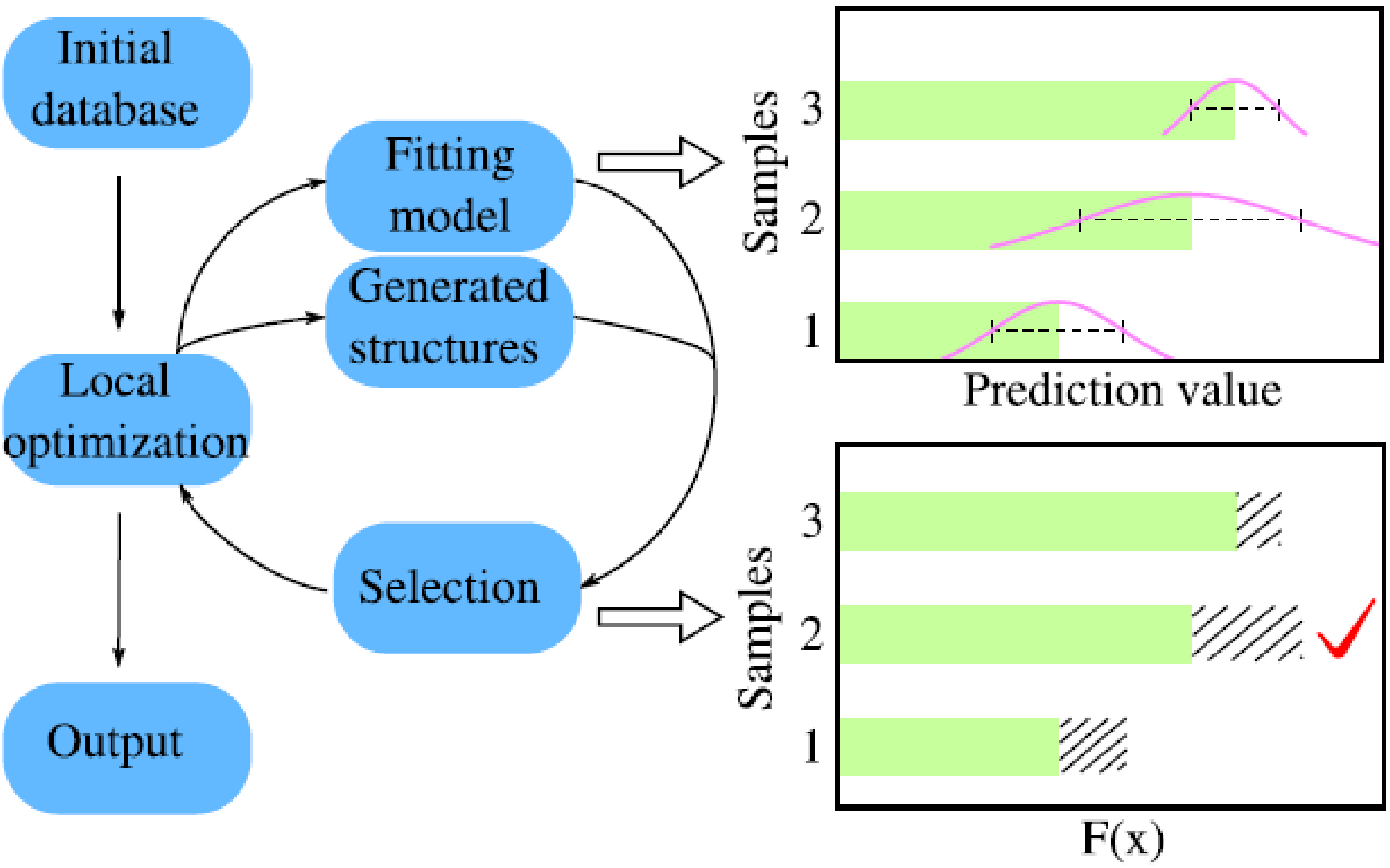

**Fig. 1** Flow chart of the machine-learning accelerated crystal structure prediction method, based on Bayesian optimization (the left panel) and the strategies of model fitting and selection (the right panel). In the upper part of right panel, the Gaussian distributions and uncertainties are depicted by pink solid curves and black dashed lines, respectively. The lower part shows the selection process by comparing the acquisition function based on the model. The estimated uncertainties are filled by black twills and the total bars represent the values of the acquisition function $F(x)$. Here the second sample is regarded as the best prediction choice (red tick) for its highest value of $F(x)$ ($k$ = 1), although the predicted average value is lower than the third sample.

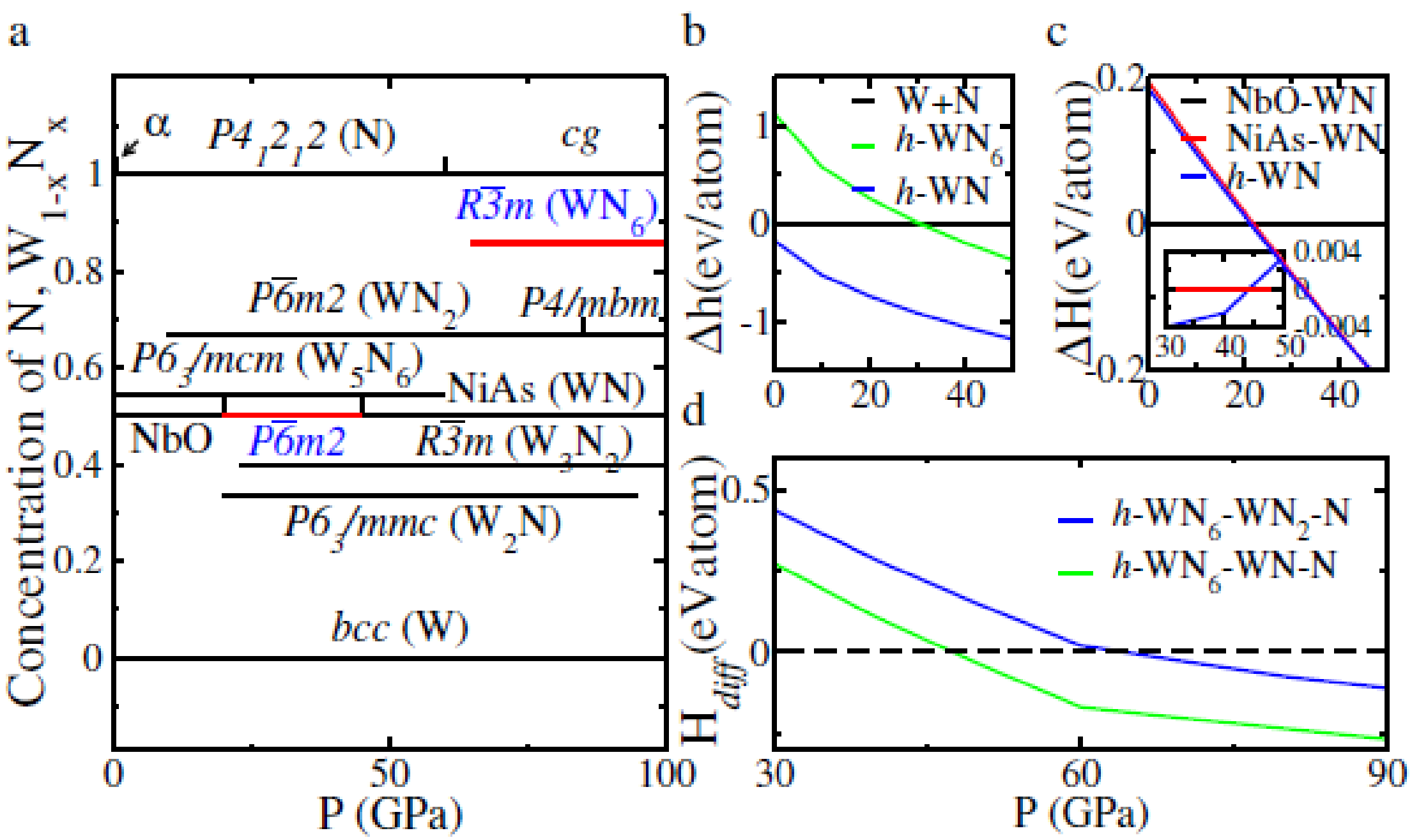

**Fig. 2** **a** The corresponding pressure-concentration diagram of stable phases. Here black ones are previously known W-N compounds. **b–d** The energy-pressure relations of $P\bar{6}m2$ WN and $R\bar{3}m$ $WN_6$. **b** The formation energy versus pressures of *h*-$WN_6$ and *h*-WN predicted in this work. Black dotted line in **d** represents the mixture of WN and N or $WN_2$ and N.

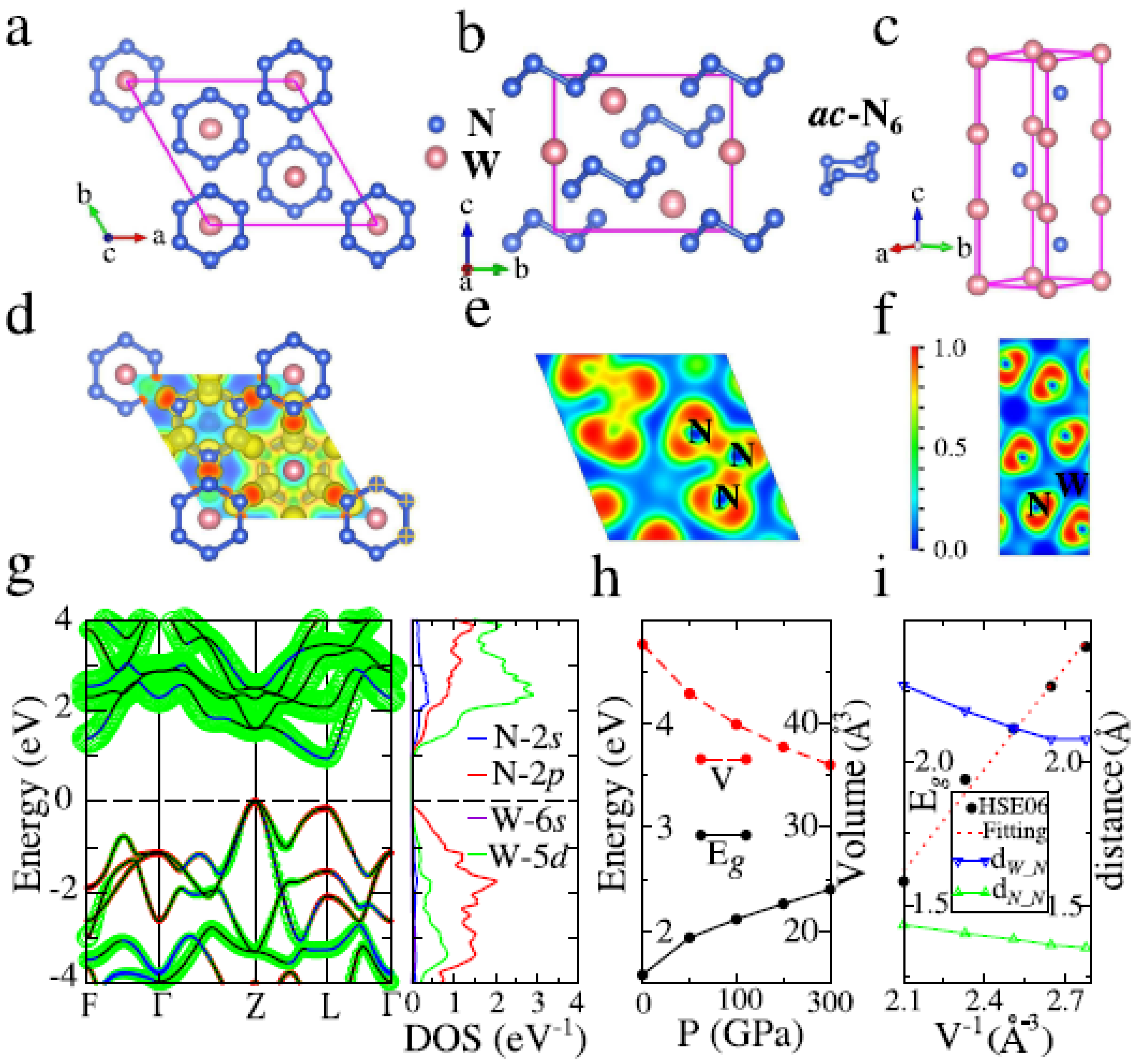

**Fig. 3 a,b and c** Crystal structures for *h*-$WN_6$ and *h*-WN, respectively; **d–f** the electron localization functions (ELF) contour plot of *h*-$WN_6$, **g** electronic band structures (left panel) and partial DOS (right panel) of *h*-$WN_6$ at 0 GPa and the HSE06 results of **h** band-gap (Eg) and equilibrium volume (V) versus pressure, and **i** Eg , $d_{W_N}$ and $d_{N_N}$ versus corresponding $V^{-1}$. The ELF contour maps of *h*-$WN_6$ are plotted in **d** with an isosurface value of 0.8 *e*/$Bohr^3$. Colored lines characterize the main contributions from atomic orbitals to bands. The zero energy is set to the top of the valence band.

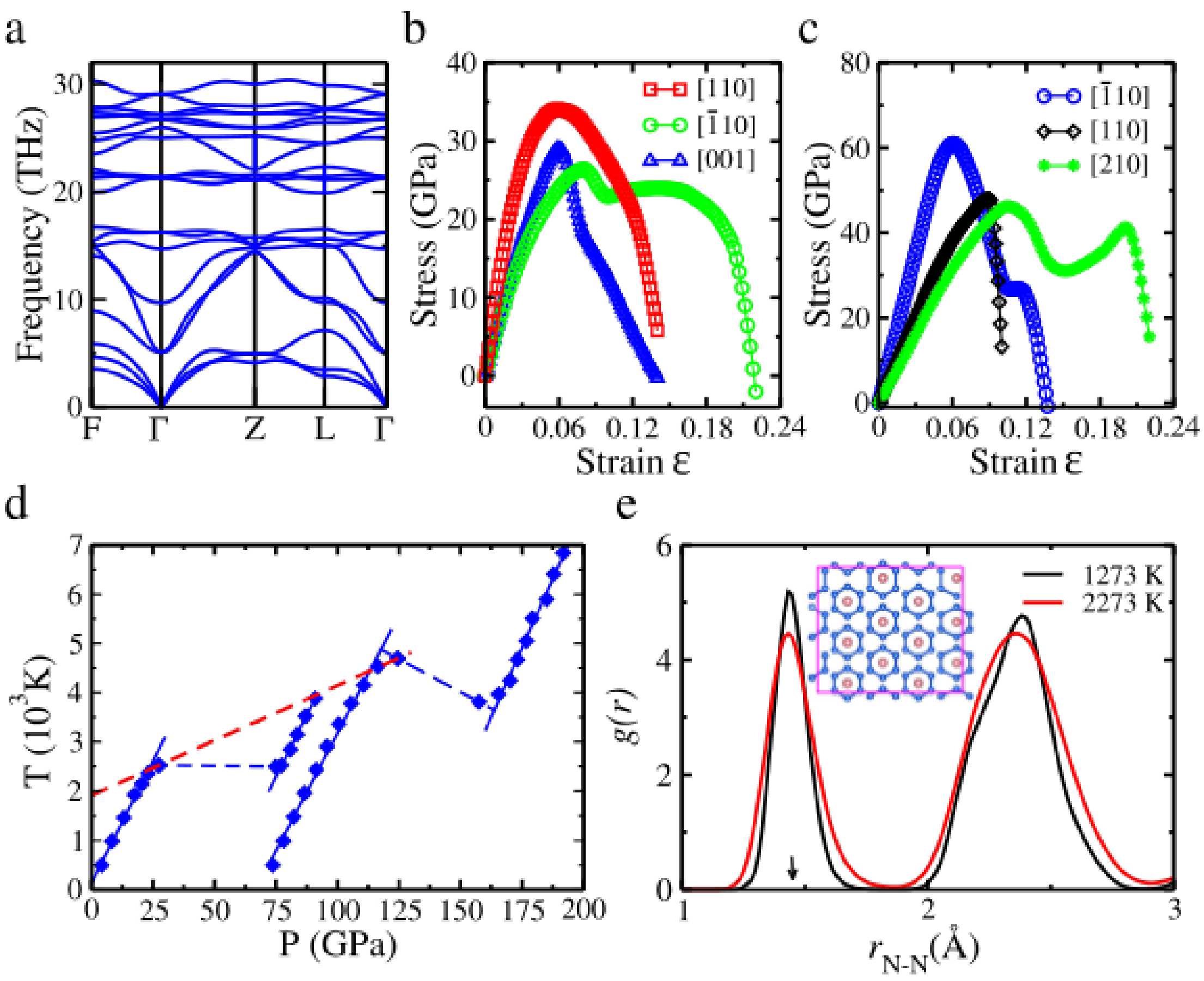

**Fig. 4 a** The phonon dispersion curves of $WN_6$ at 0 GPa, **b** the calculated stresses under the ideal tensile strains along the [$\bar{1}$10], [110], and [001] directions, **c** calculated stress responses to the Vickers indentation shear strains in the (001) plane along the [$\bar{1}$10], [110] or [210] shear direction, **d** melting temperature at ambient pressure estimated using *Z* method for *h*-$WN_6$, and **e** pair distribution functions (*g(r)*) for the N-N pairs observed during *AIMD* simulations for *h*-$WN_6$ at ambient pressure and temperatures of 1273 and 2273 K. Inset in **e** is the statistically average structure at high temperature of 2273 K. The black arrow represents the average closest N-N distance in the *ac*-$N_6$ ring. The red dashed line represents the estimated melting curve.

**Table 1** The calculated elastic constants $C_{ij}$ (GPa), bulk and shear moduli ($B$ and $G$) (GPa), and the Vickers' hardness of our $R\bar{3}m$ $h$-$WN_6$, $P\bar{6}m2$ $h$-WN and $h$-$MoN_6$ by PBE calculations at 0 GPa, in comparison with hexagonal $WN_2$ and $\delta$-MoN. Here $Hv$, $Hvc$ and $Hvt$ represent the Vickers hardness from the experiments, estimated from the methods of Chen *et al.*[54], and Tian *et al.*[36], respectively.

| Compounds | Work | $C_{11}$ | $C_{12}$ | $C_{13}$ | $C_{33}$ | $C_{44}$ | $C_{66}$ | $B$ | $G$ | $Hvc$ | $Hvt$ | $Hv$ |
|---|---|---|---|---|---|---|---|---|---|---|---|---|
| $\delta$-MoN | this work | 570.3 | 213.9 | 243.8 | 768.5 | 282.9 | 178.2 | 368.0 | 229.3 | 24.6 | 25.2 | - |
| | experiment[13] | - | - | - | - | - | - | 335 | 220 | - | - | 30 |
| $h$-$WN_6$ | this work | 662.1 | 76.1 | 132.9 | 716.5 | 359.6 | 293.0 | 302.7 | 315.7 | 57.9 | 56.8 | - |
| $h$-WN | this work | 685.9 | 208.9 | 247.3 | 721.6 | 239.0 | 238.5 | 388.9 | 236.0 | 24.2 | 24.9 | - |
| $h$-$MoN_6$ | this work | 551.4 | 68.2 | 114.1 | 625.0 | 312.2 | 241.6 | 257.9 | 268.6 | 52.3 | 50.6 | - |